**Main Manuscript for**

# Imperceptible stress sensor with interpenetrating bio-electronic interface enabled by UHMWPE nanomembrane

Yuchen FENG[c], Andreas Kenny OKTAVIUS[c], Reno Adley PRAWOTO[d], Hing Ni KO[a], Qiao GU[b], Ping GAO[ab]

[a]Advanced Materials Thrust, Function Hub, The Hong Kong University of Science and Technology (Guangzhou), Nansha District, Guangzhou, 511453, China

[b]Department of Chemical and Biological Engineering, The Hong Kong University of Science and Technology, Clear Water Bay, Hong Kong, 999077, China

[c]Advanced Materials Thrust, Interdisciplinary Office, The Hong Kong University of Science and Technology, Clear Water Bay, Hong Kong, 999077, China

[d]Microelectronics, Interdisciplinary Office, The Hong Kong University of Science and Technology, Clear Water Bay, Hong Kong, 999077, China

**Corresponding author:** Ping GAO.

**Email:** kepgao@ust.hk

**This PDF file includes:**

> Main Text
> Figures 1 to 4

## Abstract

Epidermal skin sensors have emerged as a promising approach for continuous and non-invasive monitoring of vital health signals, but to maximize their performance, these sensors must integrate seamlessly with the skin, minimizing impedance while maintaining the skin's natural protective and regulatory functions. In this study, we introduce an imperceptible sweat sensor that achieves this seamless skin integration through interpenetrating networks formed by a porous, ultra-thin, ultra-high molecular weight polyethylene (UHMWPE) nanomembrane. Upon attachment to the skin by van der Waals force, the amphiphilic sweat extrudates infuse into the interconnected nanopores inside the hydrophobic UHWMPE nanomembrane, forming "pseudo skin" nanochannels for continuous sweat perspiration. This integration is further enhanced by the osmotic pressure generated during water evaporation. Leveraging the efficient transport of biomarkers through the "skin" channels within the porous membrane, we developed an organic electrochemical transducer (OECT) cortisol sensor via in situ synthesis of a molecularly imprinted polymer (MIP) and poly(3,4-ethylenedioxythiophene) (PEDOT) within the nanomembrane, demonstrating the capability to detect cortisol concentrations from 0.05 to 0.5 µM for seamless monitoring of stress levels. This work represents a significant advancement in self-adhesive sweat sensors that offer imperceptible and real-time non-invasive health monitoring capabilities.

**Significance Statement**

This work presents an imperceptible sweat sensor utilizing UHMWPE nanomembranes. These nanomembranes seamlessly integrate with the skin, forming interconnected sweat perspiration channels for non-invasive real-time monitoring of vital health signals. This advancement in self-adhesive sweat sensors represents a significant breakthrough, offering comfortable and accurate monitoring capabilities for personalized health assessment.

**Main Text**

**Introduction**

Epidermal electronics offer a promising approach for continuous and non-invasive monitoring of vital health signals. In the field of epidermal electronics, there has been a notable shift from focusing on flexibility to prioritizing conformability and permeability after decades of development. This shift is driven by the recognition of the significance of seamlessly conforming to the skin's surface while allowing the passage of necessary substances (1, 2). Due to the seamless and stable bio-electronics interface, these electronic systems not only enhance device functionality by reducing interfacial impedance but also ensure long-term biocompatibility by minimizing mechanical discrepancies and skin irritation (3–6). Various electronic devices with these desired properties have been proposed, including substrate-free device (7–9) and porous substrate-based device (10–14).

Two questions are to be addressed for these devices to be effectively used for monitoring key biomarkers in sweat. First, a clear understanding is required of bio-electronic interface and mass transport within the permeable system. Such understanding will enable electronic designs that maximize biomarker sensitivity for seamless integration. Second, the skin's natural metabolic functions should not be hindered when affixed with these sensors. The skin functions to protect external attack factors and maintain thermal and moisture balance by self-regulated metabolic processes. The optimal design of epidermal electronics should aim to extend or enhance the natural functions of the skin. The secretion of natural electrolyte, sweat, could be utilized to reduce interfacial impedance and increase conductivity.

However, most of the current devices in the field are primarily focused on monitoring physical activity rather than capturing chemical information. These devices encounter challenges when it comes to effectively monitoring biomarkers in sweat, due to limited understanding of the bio-electronic interface and the mass transport processes within permeable systems. Moreover, many existing electronic devices hinder the inherent functions of the skin, compromising its ability to provide protection against external factors and maintain regulatory functions. To address these issues, devices that can maintain the natural function of the skin underneath whilst collecting key biomarkers from sweat seamlessly are to be developed.

In this study, we present an imperceptible sensor that incorporates a interpenetrating bio-electronics interface, preserving the skin's natural functions while enabling cortisol monitoring. The sensor utilizes an ultrahigh molecular weight polyethylene (UHMWPE) nanofibrous membrane as a "pseudo skin" - a highly versatile platform characterized by exceptional thinness, high porosity, low bending stiffness, and excellent optical transparency (15). This "pseudo skin" layer not only efficiently transports key biomarkers to the sensor, but also prevents the accumulation of salt ions on the skin surface. Building upon this, we fabricated a multilayer stack skin sensor by in situ synthesis of molecularly imprinted polymer (MIP) and poly(3,4-ethylenedioxythiophene) (PEDOT) on the nanomembrane. The resulting organic electrochemical transistor (OECT) sensor exhibits selective absorption, conductivity, and the ability to detect sweat cortisol within the range of 0.05 to 0.5 µM, all while preserving its imperceptible and biocompatible characteristics.

**Results and Discussions**

The skin functions as a permeable barrier, providing protection against microorganisms and harmful compounds, and regulates body temperature through the moisture balance (16). When placing a 300 nm-thick UHMWPE nanofibrous membrane onto the skin, it forms an interpenetrating interface, maintaining the essential attribute of a permeable barrier, resulting in a "***pseudo skin***" (**Figure 1a**).

The membrane was first attached to the skin surface via van der Waals forces. This attachment was enhanced by ethanol spraying through the pores in the membrane, which was then followed by ethanol evaporation from the membrane pores. This process removes the air gap between the membrane and the skin surface, strengthening the van der Waals forces, **Figure 1b**. Subsequently, the interconnected nanopores bounded by the ultrathin fibers within the nanomembrane enable the infusion of sweat extrudates, forming a pseudo skin-like sweat channel, **Figure 1c**.

The interpenetrating interface transports sweat through pseudo skin-like channels. The osmotic pressure created by the evaporation of sweat on the membrane's exterior surface maintains the integrity of the interface. This, in turn, enhances the durability and stability of the interface. The ToF-SIMS (Time-of-Flight Secondary Ion Mass Spectrometry) results, obtained on the nanomembrane peeled off from the skin, (**Figure 1d, f, Figure S1, Table S4**) provide insights into the spatial distribution of the stratum corneum and sweat secretion compounds along the thickness of the membrane (17–20), confirming the existence of the interpenetrating interface and its resulting properties. Physically, the presence of the stratum corneum retained inside the membrane is observed by comparing the morphology changes between the pristine and the delaminated membrane from the epidermis - the pores in the pristine membrane are visibly blocked by non-vaporizable sweat extrudates, or the stratum corneum (**Figure 1e**). We believe the stratum corneum-filled membrane channels may serve as a pseudo-epidermal skin layer, enabling seamless sweat transportation from the skin to the atmosphere (**Figure 1c**).

The spatial concentration profile of chloride ions, known for their high mobility in aqueous solutions, offers insights into the sweat transport process (**Figure 1d**). Initially, most of chloride ions are transported along with the sweat and distributed within the membrane. As water evaporates into the atmosphere, the chloride ions become concentrated on the top surface of the membrane in contact with the ambient. This higher concentration of chloride ions on the exterior surface results in a large osmotic pressure, which acts to both pump the ions out of the membrane channels and maintain stronger attachment of the membrane to the skin. This observation provides evidence that the interpenetrating interface between the membrane surface and the stratum corneum enables biomarkers to permeate the membrane and reach the sensor, while preventing salt accumulation on the skin.

The utilization of the UHMWPE membrane in biosensors offers several advantageous features attributable to this interpenetrating interface. During the resting stage, the skin secretes insensible sweat at a rate of approximately 30 g/m$^2$/h, and the membrane allows for the transportation of 97% of water vapor, ensuring efficient moisture transfer. Under sweating conditions, where the back and forehead may exhibit sweat rates exceeding 700 g/m$^2$/h, the membrane demonstrates a remarkable water vapor transfer rate (WVTR) of up to 612.5 g/m$^2$/h (**Figure S2**).

The UHMWPE membrane exhibits a contact angle of approximately 120° with water, which increases the hydrophobicity of the skin (**Figure S3**). This property effectively enhances the barrier function of the skin by hindering the infiltration of contaminants and viruses when exposed to the surrounding air. Additionally, this hydrophobic property in conjunction with the pseudo skin-like characteristic of the membrane ensures no cross-contamination of the device. This is evidenced by the absence of sweat water droplet accumulation on the membrane surface, in contrast to the natural skin (**Figure 1b**).

In the context of conformability, the interface energy can be expressed as $U = U_{skin} + U_{bending} - |U_{adhesion}| < 0$, where $U_{skin}$, $U_{bending}$ and $U_{adhesion}$ represent the elastic energy of the skin, the bending energy of the materials and the interfacial adhesion energy, respectively (21). Achieving conformability relies on a combination of low flexural rigidity of materials and high adhesion. The flexural rigidity of a film is defined as $Et^3/12(1-v^2)$, where E, t, and v represent Young's modulus, thickness, and Poisson's ratio, respectively. In the case of nanofibrous materials adhering to the skin primarily through van der Waals forces, their thickness and the contact area between the nanofibers and the skin play crucial roles (22, 23). Therefore, reducing thickness and increasing the surface area are the most effective approaches to achieving conformability. In this study, the UHMWPE membrane used as a substrate shows a uniform thickness of approximately 300 nm, equivalent to a solid film thickness of 142 nm, as confirmed by atomic force microscopy (AFM) characterization (**Figure S4**). To achieve high skin adhesion, the flexural rigidity of a polymeric film should be below $10^{-2}$ nN*m, which falls within the range observed in living brain slices (24). The flexural rigidity is calculated to be $2.09 \times 10^{-3}$ nN*m. As a result, the membrane-coated artificial skins exhibited surface roughness features that closely resemble those of bare artificial skins. Microscopic observations consistently reveal the preservation of peaks and valleys on the curvilinear surface of the membrane-covered artificial skins (**Figure S5**). Even when subjected to biaxial stretching (**Figure S6**), the nanofibrous membrane demonstrated exceptional conformability and maintained intimate contact with the curvilinear surface. These results prove the exceptional self-adhesiveness of UHMWPE membrane, which does not necessitate surface modification or additional adhesive (**Figure S7**). Overall, the UHMWPE membrane serves as an imperceptible pseudo skin, demonstrating excellent capabilities in terms of both conformability and permeability (**Figure S8**).

To ensure the pseudo skin-like and imperceptibility characteristic of the like UHMWPE nanomembrane are retained after being fabricated into a sensor device, we first functionalized the UHMWPE nanomembrane by in situ synthesis and synthesized MIP@UHWMPE and PEDOT@PVA-coated PE, respectively.

The MIP@UHMWPE composite membrane was created using the widely employed surface imprinting method, which is known for its effectiveness in nanoparticle synthesis (25, 26). The inclusion of MIP in the sensor design was intended to leverage its selective and rapid rebinding capability for target species, e.g., cortisol following the removal of target species after synthesis. Specifically, we devised a Janus membrane by synthesizing a monolayer of MIP nanoparticles on one side of the membrane surface while leaving the other side intact (**Figure 2a**). This design takes advantage of the unique skin-like property of the PE nanomembrane to interface with skin, allowing cortisol molecules along with sweat extrudates to enter the interconnected porous channels on the nascent membrane side facing the skin, where they become adsorbed onto the MIP nanoparticle surfaces on the opposite side.

To ensure consistent attachment of MIP nanoparticles onto the surface of the membrane, we implemented the following procedure: initiators were initially loaded into the pores of the UHMWPE nanomembrane. Subsequently, the membrane loaded with initiators was placed onto the surface of a reservoir containing solutions of monomer with/without cortisol target for interfacial synthesis of MIP/or NIP nanoparticles. The hydrophobic nature and small pore sizes of the UHMWPE nanomembrane ensured that the growth of MIP nanoparticles occurred exclusively on the nanomembrane surface facing the monomer. The entire system was then exposed to 365 nm UV light, irradiating the membrane's external surface. This UV light initiated the polymerization of monomers and facilitated the decomposition of initiators in the vicinity of the membrane surface. Over time, these polymers developed into nanoparticles, which were separated from the solvent due to their limited solubility. The growth of these nanoparticles continued on the nanofibrillar network, as depicted in **Figure S9**. After an hour of reaction time, well-dispersed nanoparticles with an average diameter of approximately 150 nm were formed, while the opposite side of the membrane remained unaffected (**Figure 2b-d, g and Figure S10-12**). To eliminate any loosely bound nanoparticles or unreacted monomers, the membranes underwent rigorous ultrasonication. The resulting MIP@UHMWPE membrane, as shown in **Figure 2c-d**, displayed a Janus structure, with one surface uniformly coated by a near-monolayer of MIP nanoparticles and the other surface retaining the original PE structure.

The synthesized MIP nanoparticles in **Figure S12** comprise methyl acrylic acid (MAA) and ethylene glycol methacrylate (EGDMA) after the removal of the template compound cortisol. The successful synthesis of the MIP@PE composite membrane is confirmed by the presence of a strong absorption peak at 1730 cm$^{-1}$, corresponding to the C=O stretching vibration, as well as peaks at 1255 and 1150 cm$^{-1}$ attributed to the C-O stretching vibration (**Figure 2e**). These findings are supported by the Raman spectra (**Figure 2f**). To demonstrate its high sensitivity, the MIP@PE composite membrane was immersed in cortisol solutions with concentrations as low as 50 ppm. The weight uptake of cortisol was quantified using high-performance liquid chromatography (HPLC) (**Figure S14**). As shown in **Figure S15**, the MIP particles exhibit significant adsorption performance after 180 minutes, with a continuous increase in absorption capacity over time. The MIP@PE composite membrane achieves a maximum adsorption capacity of 591.26 µg/g, highlighting its potential for cortisol detection and monitoring applications.

To maintain the thin and porous characteristics of the UHMWPE membrane after converting it into a semiconductor nanomembrane, we developed a vapor phase polymerization method to synthesize PEDOT (a commonly used organic semiconductor in OECTs) directly onto the nanofibril surfaces within the membrane. This process resulted in a porous yet consistently conductive composite membrane (**Figure 3a**). Our synthesis protocol began with the preparation of a hydrophilic substrate by coating the PE membrane with PVA (**Figure 3b and Figure S16-17**), ensuring uniform catalyst distribution in the subsequent phase. Using a meniscus-guided coating method, which leverages the interface curve formed by a liquid on a solid surface, we uniformly deposited a layer of catalyst (FeCl$_3$) onto the membrane surface, achieving precise and homogeneous coating (27). We then introduced the monomer in gaseous form into the reaction chamber, where chemical oxidative polymerization occurred at the interface between the catalyst and monomer, facilitating the formation of the PEDOT polymer within the UHMWPE scaffold. After the reaction, we eliminated any excess catalyst and unreacted monomer, resulting in the formation of the PEDOT@PVA-PE membrane (**Figure 3c-d**). This approach allowed us to create a composite membrane that preserves the desirable structural properties of the UHMWPE while incorporating the conductive properties of PEDOT.

We conducted FTIR and micro-Raman spectroscopic characterization of the PEDOT@PE membrane to confirm its successful synthesis. FTIR spectroscopy (**Figure 3e**) revealed characteristic peaks indicative of PEDOT's molecular structure: C-S-C stretching vibrations at 979 cm$^{-1}$, 919 cm$^{-1}$, 836 cm$^{-1}$, and 687 cm$^{-1}$, and stretching vibrations of C=C and C-C at 1523 cm$^{-1}$ and 1346 cm$^{-1}$, respectively; and C-O-C stretching vibration at 1199 cm$^{-1}$, 1141 cm$^{-1}$, 1087 cm$^{-1}$, and 1051 cm$^{-1}$. The visual transformation of the PEDOT membrane from transparent to light blue indicates its synthesis and suggests partial porosity while maintaining a continuous phase, which is crucial for maintaining conductivity. Raman spectroscopy was employed to evaluate the oxidized state and doping level of the PEDOT membrane. Deconvolution of the symmetric C=C stretching peak at 1440 cm$^{-1}$ revealed two distinct components: one at 1408 cm$^{-1}$ associated with the quinoid form and another at 1443 cm$^{-1}$ linked to the benzenoid form (**Figure 3f and Figure S18**) (27, 28). These peaks, commonly used to estimate the oxidized state of PEDOT, indicate a higher ratio of the quinoid form in the synthesized PEDOT@PE composite membrane, contributing to its enhanced conductivity, which measures 11.36 S/cm.

Cortisol, a steroid hormone, released into biofluids during psychological or physiological stress, serves as a biomarker for various stress-related diseases, including cardiovascular disorders and weakened immune responses. (29–32). To detect variations of cortisol levels in sweat, we designed a sweat sensor based on our UHMWPE nanomembrane. The designed sensor comprises two key components: an OECT acting as the transducer and a MIP layer serving as the selective sensing element. The overall device design is illustrated schematically in Figure 4b, where the MIP@PE and PEDOT@PE layers were synthesized in situ following the methods described in the previous sections (**Figure S19-21**). Importantly, the stack structure's porous nature and thinness ensure maintained sweat permeability (**Figure S2**), a critical feature for accurate and responsive cortisol detection.

The SEM cross-section analysis of the fabricated OECT provides insights into its layered structure. The topmost layer is composed of a Pt thin film deposited on the PE fiber using the PVD technique, forming three electrodes (source, drain, and gate). Below this layer, there is a porous yet continuously conductive composite of PEDOT@PE, as shown in **Figure S22-23**. The microstructure analysis indicates that the drain and source electrodes are exclusively connected by PEDOT, and no instances of potential short-circuiting have been detected. The spatially connected structure of the OECT is further supported by depth profile analysis and surface images obtained for ToF-SIMS results, as shown in **Figure S24**.

The UHMWPE-based OECT described in this study could be operated in both depletion mode and accumulation mode, where a hydrogel swollen with 0.9 wt% NaCl aqueous solution was employed as the electrolyte. In the accumulation mode, a negative gate voltage is applied, causing anions to migrate through the electrolyte and penetrate the PEDOT channel. Consequently, the PEDOT material undergoes doping and the accumulation of positively charged holes within the channel, resulting in an increase in drain current ($I_D$). Similarly, in the depletion mode, the drain current decreases due to a reduction in the number of charge carriers (**Figure 4c**). The UHMWPE-based OECT device exhibits a transconductance value of 18 µS at a gate voltage ($V_G$) of -0.2V and a drain voltage ($V_D$) of 0.95V (**Figure 4d**), which serves as the operational parameter for the sensor. **Figure S25** demonstrates the satisfactory stability and data reproducibility of the UHMWPE-based OECT by showing the variation of ID in response to the change in VG within the range of -0.5V to -0.1V.

Through the integration of the UHMWPE-based OECT and the MIP@UHMWPE composite membrane, we have developed an UHMWPE-based sweat sensor for the purpose of detecting cortisol levels in sweat. In the Janus MIP@UHMWPE configuration, the PE fiber faces the skin, creating an interpenetrating interface. Cortisol molecules carried by sweat vertically migrate from the MIP particle layer to the PEDOT layer within this interface. The MIP nanoparticles contain pores that act as ion channels for sweat electrolytes. When cortisol molecules selectively bind to the MIP nanoparticles via hydrogen bonds, they block the ion channels, causing a decrease in ionic current. This reduction in current reliably indicates the presence of cortisol. (**Figure 4b**). This mechanism has been extensively validated and widely utilized in the field of sweat sensors (33–35).

The experimental results demonstrate a gradual decrease in current corresponding to increasing levels of cortisol, where 0.9 wt% NaCl and different aqueous solutions with varying cortisol concentrations is utilized (**Figure 4e**). Notably, while the sensor saturates beyond 1 µM, it exhibits excellent calibration within the range of 0.05 µM to 1 µM, with an R-square value of 0.98 (**Figure 4f**). During on-body testing, the UHMWPE-based sweat sensor is directly applied onto the skin for cortisol monitoring (**Figure 4g and h**). An additional UHMWPE membrane covering it as an insulating layer. This breathable and hydrophobic membrane enables water evaporation while preventing non-localized analytes from interfering with the sensor's performance. This sensor design not only ensures comfort when attached to the skin but also enhances the accuracy of cortisol detection by minimizing interference from other substances present in sweat. Following a one-minute application of the cold pressor test (CPT), a noticeable increase in cortisol concentration is observed within the sweat, which is subsequently detected and monitored by the integrated sensor.

**Conclusion**

In conclusion, this study presented an imperceptible sweat sensor based on a UHMWPE nanofibrous membrane. The nanomembrane exhibited a pseudo skin-like character, including an interpenetrating interface and highly interconnected nanopores, which preserved the natural protective, temperature-regulating, and sweat permeation functions of the underlying skin. Sweat extrudates integrated into the membrane's pores, creating continuous perspiration channels. The self-supporting, self-adhesive, and durable nature of the nanomembrane allowed for in-depth characterization using ToF-SIMS analysis, revealing the distinctive interface and mechanism of biomarker transport within the sensor.

Through the application of customized in-situ synthesis methods, the membrane is skillfully functionalized to possess both selectivity and conductivity. These composite materials (MIP@UHMWPE and PEDOT@UHMWPE) successfully retain the thin and porous characters, preserving self-adhesive and breathable properties of the resulting UHMWPE-based sweat sensor. This technological advancement enables seamless monitoring of stress levels for personal health applications. Importantly, it shows immense potential in overcoming challenges associated with prolonged wear and facilitating the vertical integration of self-adhesive sweat sensor devices. Consequently, this innovation not only addresses potential concerns regarding skin health but also paves the way for practical utilization of these sensors in real-world scenarios.

**Materials and Methods**

**Materials**: Poly(vinyl alcohol) (PVA, P8136), hydrogen chloride (HCl), ethylene glycol dimethacrylate (EGDMA), 2-methacrylic acid (MAA), azobisisobutyronitrile (AIBN), hydrocortisone (cortisol) and 3,4-Ethylenedioxythiophene (EDOT) were purchased from Sigma-Aldrich. UV curable hydrogel (#JN0917-A) was purchased from Polychem UV/EB International Corp. The ultrahigh molecular weight polyethylene (UHMWPE) membrane was prepared using the patented method developed by our group. Porous and hand-manipulatable UHMWPE membranes with thicknesses of 300 nm that show full conformability to the skin and can be detached for biomarker analysis after biomarker collection from sweat are used as the substrate in this study.

**Characterization**: Scanning electron microscopy (SEM) was used to characterize the surface morphology of the membranes. The samples were sputter-coated using a thin layer of platinum to minimize surface charging and was characterized by a field emission scanning electron microscopy (JOEL, JSM-7800F). The cross-section of samples was characterized by Dual Beam FIB/SEM (FEI Helios G4 UX), in which the stage has a tilt angle of 52°. The chemical structure of membranes was characterized by Fourier-transform infrared spectrometry (FTIR) (Thermo Fisher Nicolet, iS50), Raman spectrometry (Renishaw, InVia) and X-ray photoelectron spectroscopy (Axis Ultra DLD). All FTIR spectra were collected in the transmitted mode with an accumulation of 32 scans at a resolution of 4 cm-1. The laser light excited with the 514 nm line was focused on the sample with a microscope objective (Leica, 50x magnification) when measuring Raman spectra. Conductivity is measured by Resistivity measurement system (Bio-Rad, HL5500P). Air permeability tester (RH-TQG645, Guangzhou Runhu Instruments Co., Ltd) is employed to measure air permeability of membrane. The composition of the solution was assessed using high-performance liquid chromatography (HPLC) (Waters 2695) and ion chromatography (IC) (Metrohm/881 IC). Sweat samples were collected utilizing the Macroduct sweat collection system (Model 3700 SYS).

**Preparation of MIP@UHMWPE**: The MIP@PE membrane was prepared by in situ synthesis of MIP nanoparticles on the UHMWPE membrane. Photo-polymerization was utilized to synthesized MIP nanoparticles, which allows low synthesis temperature and thus avoids breakage of hydrogen bond. The UHMWPE membrane was immersed inside MIP precursor solution (monomer concentration = 8.0 wt.%, template:functional monomer:crosslinker = 1:4:30, initiator = 2.0 wt.% of monomer). The photosynthesis was conducted under the UV irradiation (365 nm, 10 mW/cm2) at 4℃. After the completion of the polymerization, the synthesized MIP@PE membrane was washed by rinsing with pure ethanol under sonication to remove cortisol, unreacted monomers, and excess MIP particles that were not strongly confined to the membrane substrate. Non-imprinted polymers (NIP) membrane was also prepared as a control sample using a similar protocol to that for the MIP synthesis but without the use of a template.

**Preparation of PEDOT@PVA-coated PE:** The synthesis of PEDOT@PE composite membrane mainly consisted of two processes: catalyst coating and vapor phase polymerization. During catalyst coating, the PE membrane was firstly coated with PVA by dip coating to enhance its hydrophilicity. Attachment of PVA on the PE fibril surface is stabilized by the hydrophobic-hydrophobic interactions between PVA and the large specific surface areas of the nanofibrils on the PE substrate. As a result, the contact angle of this PVA-PE membrane with water decreased from ~130° to ~60° (**Figure S9**). The exposed hydroxyl groups of the PVA coating on the PE surface enhanced water evaporation and helped catalyst coating uniformity. After PVA treatment, the catalyst, $FeCl_3$ was deposited by evaporating 3 M $FeCl_3$ aqueous solution on the membrane. Then, the resulting $FeCl_3$@PVA-PE membrane was used to catalyze the vapor phase polymerization of EDOT for the synthesis of PEDOT. Briefly, EDOT and HCl vapors were generated by vacuum at 70℃. Firstly, 1M HCl in butanol was added into the system to liberate $Fe^{3+}$, which was the key for oxidation. Then, pure EDOT was added and the generated EDOT vapor was polymerized on the membrane. After that, 0.5 ml of 1M HCl in butanol was introduced as a dopant to the synthesized PEDOT to achieve higher conductivity.

**Preparation of UHMWPE-based sweat sensor:** As illustrated in Figure S19, the fabrication process of the sweat sensor based on UHMWPE membrane involves a sequential deposition of Pt, PEDOT, and MIP through a series of steps. Firstly, a shadow mask with an OECT pattern is prepared by laser cutting on cardboard and placed onto the UHMWPE membrane for PVD. This process transfer OECT pattern from shadow mask to UHMWPE membrane, generating a Pt thin film on the top layer. In the second step, a $FeCl_3$@PVA-PE membrane (4 mm x 2.5 mm) is positioned on the back side of the UHMWPE-based OECT, specifically on the side without Pt. Simultaneously, EDOT vapor is introduced to trigger the localized synthesis of PEDOT. Once the PEDOT synthesis is finished, the $FeCl_3$@PVA-PE structure is carefully peeled off, resulting in the formation of the PEDOT@PE network between the source and drain electrodes. In the final step, a MIP@PE membrane (6.5 mm x 5 mm) is integrated into the sensor assembly, completing the fabrication of the UHMWPE-based sweat sensor. To establish seamless connectivity between the fabricated sweat sensor and an external electrochemical station, a flexible cable composed of copper and commercial LMWPE is utilized. The connection between the flexible cable and the sweat sensor is achieved by ultrasonic welding.

**HPLC chromatography:** The separations were conducted using a reversed-phase high-performance liquid chromatography (HPLC) column (ACQUITY PRM BEHC18 1.7μ 2.1x150mm) employing an isocratic elution method with acetonitrile;water 6:4 (v/v) as the mobile phase at a flow rate of 0.3 mL/min. The chromatographic run was completed within 11 minutes. Detection was carried out at a wavelength of 245 nm using a photodiode array detector (PDA). The peak observed at a retention time of 1.999 minutes corresponds to cortisol (**Figure S14**). The constructed standard curve for cortisol demonstrates a high degree of linearity, as indicated by an $R^2$ value of 0.999, over the concentration range of 0.25 ppm to 100 ppm.

**Adhesion strength measurement**: The adhesion of a UHMWPE membrane to a substrate (3M, Nexcare™ Hydrocolloid Dressing HCFC-2) is facilitated by the formation of Van der Waals' force between the two materials. Initially, ethanol is sprayed onto the substrate, filling the interface between the substrate and the membrane. As the ethanol evaporates through the membrane pores, Laplace pressure develops, resulting the removal of the air trapped between the membrane and the substrate, leading to the establishment of von der Waals forces therein. The strength of the adhesion is quantified using a 180° peeling test, wherein the substrate with the membrane is placed between two parallel fixtures and peeled off at a rate of 0.1 mm/s. The results of the peeling test for samples of varying thickness and width are presented in **Figure S7**. Each condition is tested three times, and only the data in the stable region of displacement (ranging from 15 to 45 mm) is analyzed to obtain the average force and interfacial toughness (as inserted). The mean force is found to be proportional to sample width, and the interfacial toughness of samples with widths of 1 cm and 1.5 cm is observed to be very similar.

**Calculation of flexural rigidity of UHMWPE nanofibrous membrane:** Flexural rigidity is defined as

$$D = \frac{Et^3}{12(1-v^2)}$$

where E, t, and v respectively represent Young's modulus, thickness and Poisson's rate. Young's modulus of UHMWPE calculated from stress-strain curve is 7.29±0.073 GPa (**Figure S26**). Area density of UHMWPE is 0.01425 mg/cm$^2$, so its solid content is equal to 142.5 nm thick. The Poisson's rate of UHMWPE is 0.4 (5, 6). Therefore, the flexural rigidity is calculated to be 2.09 ×10$^{-3}$ nN*m.

**Acknowledgments**
The authors gratefully acknowledge financial support from Guangzhou Municipal Government. The authors gratefully acknowledge technical support from the Materials Characterization and Preparation Facility (CWB & Guangzhou). The authors gratefully acknowledge technical and financial support from Project on Preparation and industrialization of high-strength nanofibre porous membrane of ultra-high molecular weight polyethylene (UHMWPE), Research Institute of Tsinghua, Pearl River Delta with grant number RITPRD23EG01, and Guangdong Provincial Department of Science and Technology with grant number 2021JC02C152.

29. J. P. Herman, *et al.*, Regulation of the hypothalamic-pituitary- adrenocortical stress response. *Compr. Physiol.* **6**, 603–621 (2016).

30. D. Y. Lee, E. Kim, M. H. Choi, Technical and clinical aspects of cortisol as a biochemical marker of chronic stress. *BMB Rep.* **48**, 209–216 (2015).

31. L. Manenschijn, J. W. Koper, S. W. J. Lamberts, E. F. C. Van Rossum, Evaluation of a method to measure long term cortisol levels. *Steroids* **76**, 1032–1036 (2011).

32. J. Smyth, M. Zawadzki, W. Gerin, Stress and disease: A structural and functional analysis. *Soc. Personal. Psychol. Compass* **7**, 217–227 (2013).

33. D. Mukasa, *et al.*, A Computationally Assisted Approach for Designing Wearable Biosensors toward Non-Invasive Personalized Molecular Analysis. *Adv. Mater.* **35** (2023).

34. M. Caldara, *et al.*, Dipstick Sensor Based on Molecularly Imprinted Polymer-Coated Screen-Printed Electrodes for the Single-Shot Detection of Glucose in Urine Samples—From Fundamental Study toward Point-of-Care Application. *Adv. Mater. Interfaces* **10** (2023).

35. O. Parlak, S. T. Keene, A. Marais, V. F. Curto, A. Salleo, Molecularly selective nanoporous membrane-based wearable organic electrochemical device for noninvasive cortisol sensing. *Sci. Adv.* **4** (2018).
11

**Figures**

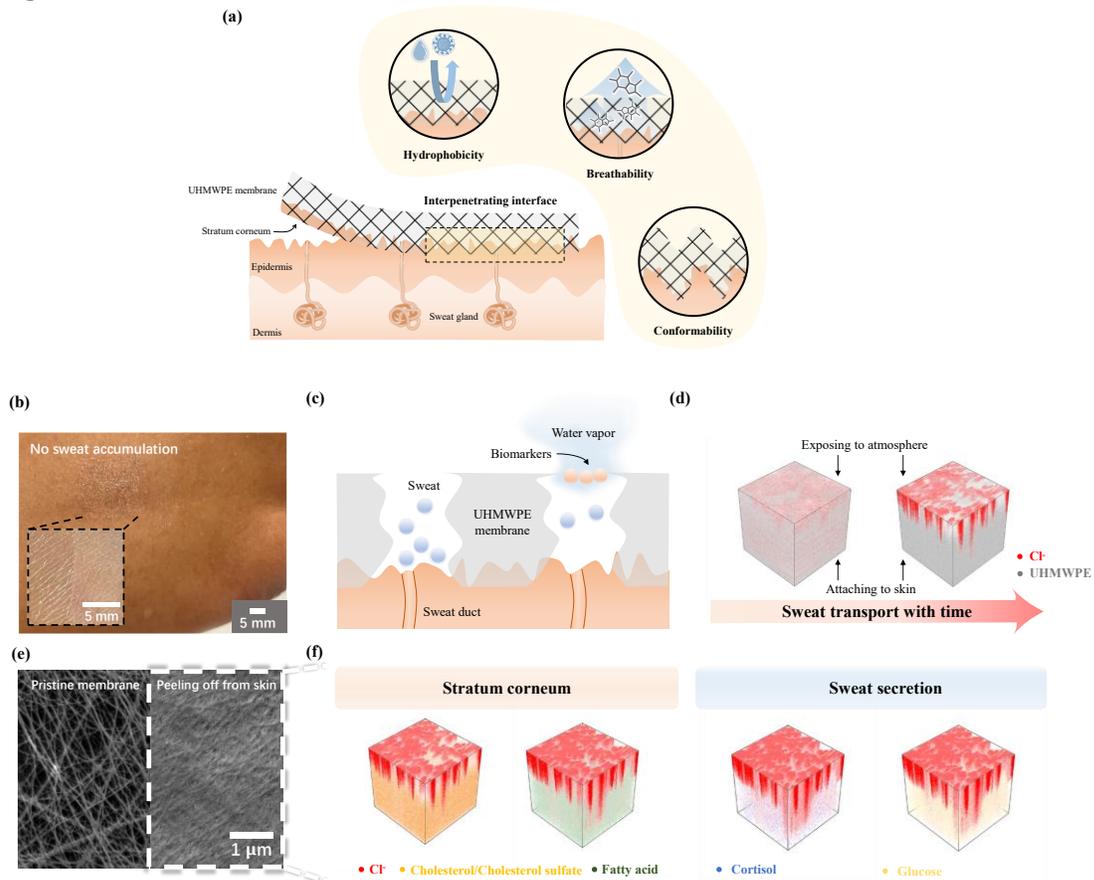

**Figure 1. The integrated interface of UHMWPE membrane and skin.** (a) schematic illustration of integrated interface between UHMWPE membrane and epidermis and its resulting properties. (b) A photograph of UHMWPE nanofibrous membrane conformed on the hand of a volunteer. (c) Schematic illustration of sweat transport in UHMWPE membrane. (d) ToF-SIMS 3D rendering of chloride and UHMWPE membrane showing the accumulation of chloride ions on the exterior surface of the membrane exposed to air due to water evaporation. (e) SEM images of membrane lateral surfaces: pristine membrane and membrane after peeling off from skin. (f) ToF-SIMS 3D distribution of sweat secretion components and sweat secretion.

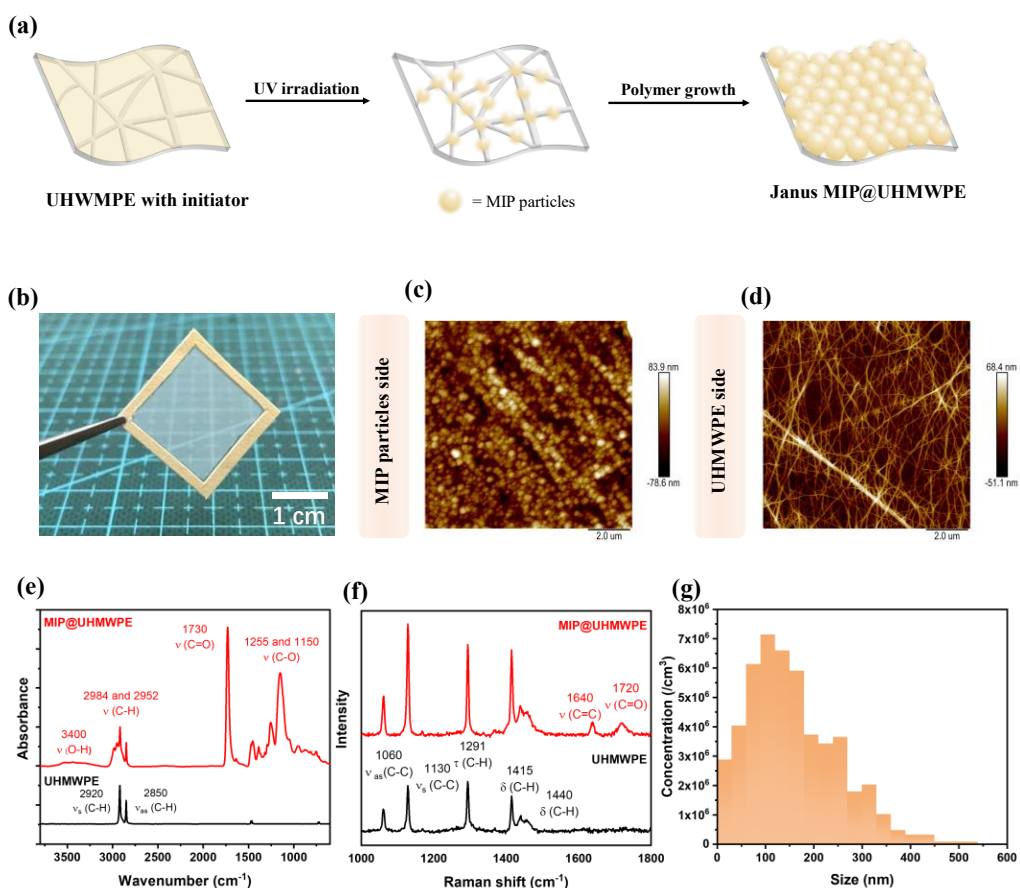

**Figure 2. In-situ synthesis of Janus MIP@UHMWPE.** (a) Schematic illustration of in-situ synthesis of MIP@UHMWPE. (b) A photograph of Janus MIP@UHMWPE membrane (c-d) AFM pictures of MIP particles side and UHMWPE side of Janus MIP@UHMWPE (e-f) FTIR and Raman spectrum of MIP@PE before and after in-situ synthesis. All FTIR spectra were normalized using the peak at 2920 cm$^{-1}$, which corresponds to the C-H symmetric stretching vibration band of PE. (g) MIP nanoparticles size distribution.

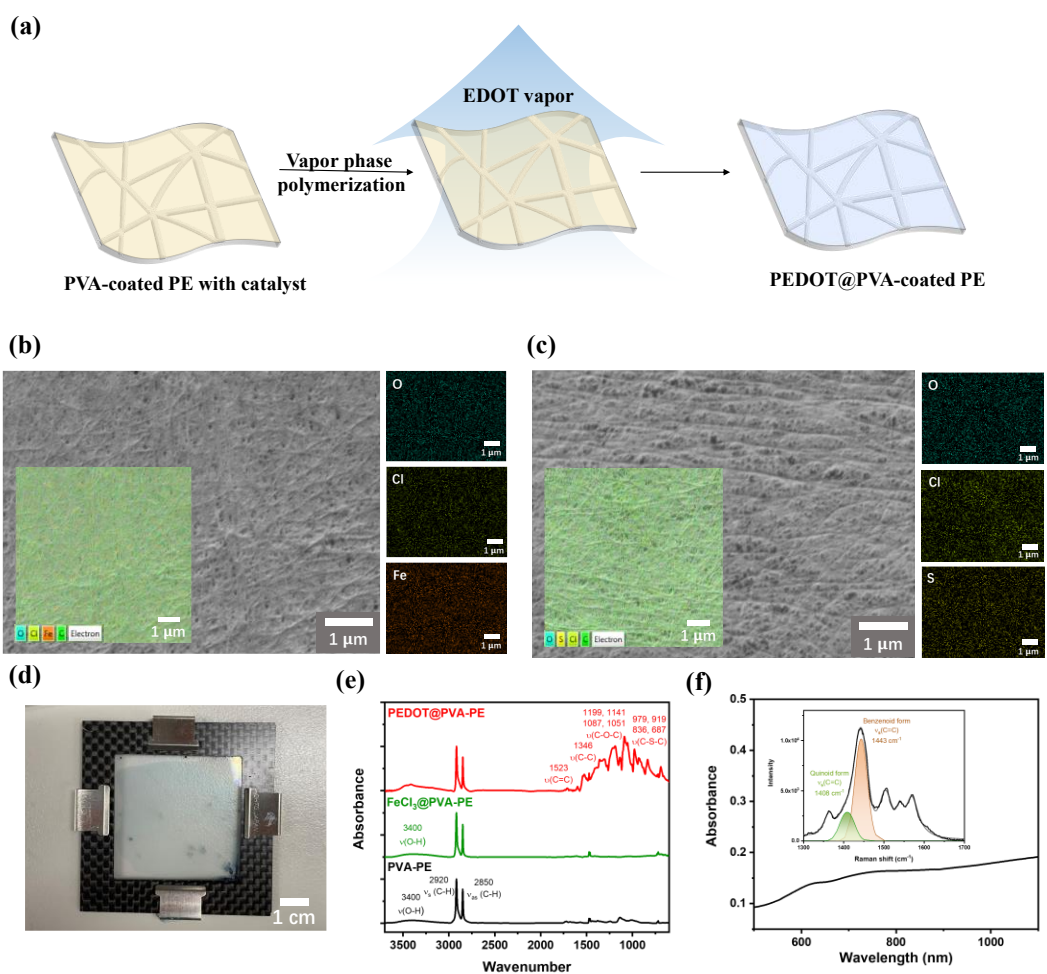

**Figure 3. In-situ synthesis of PEDOT@PVA-coated PE.** (a) Schematic illustration of in-situ synthesis of PEDOT@PVA-PE. SEM pictures and EDS result of (b) FeCl$_3$@PVA-PE and (c) PEDOT@PVA-PE. (d) A photograph of PEDOT@PVA-PE membrane. (e) FTIR spectrum of PVA-PE, FeCl$_3$@PVA-PE and PEDOT@PVA-PE. All FTIR spectra were normalized using the peak at 2920 cm$^{-1}$, which corresponds to the C-H symmetric stretching vibration band of PE. (f) UV-Vis and Raman spectrum of PEDOT@PVA-PE (inset).

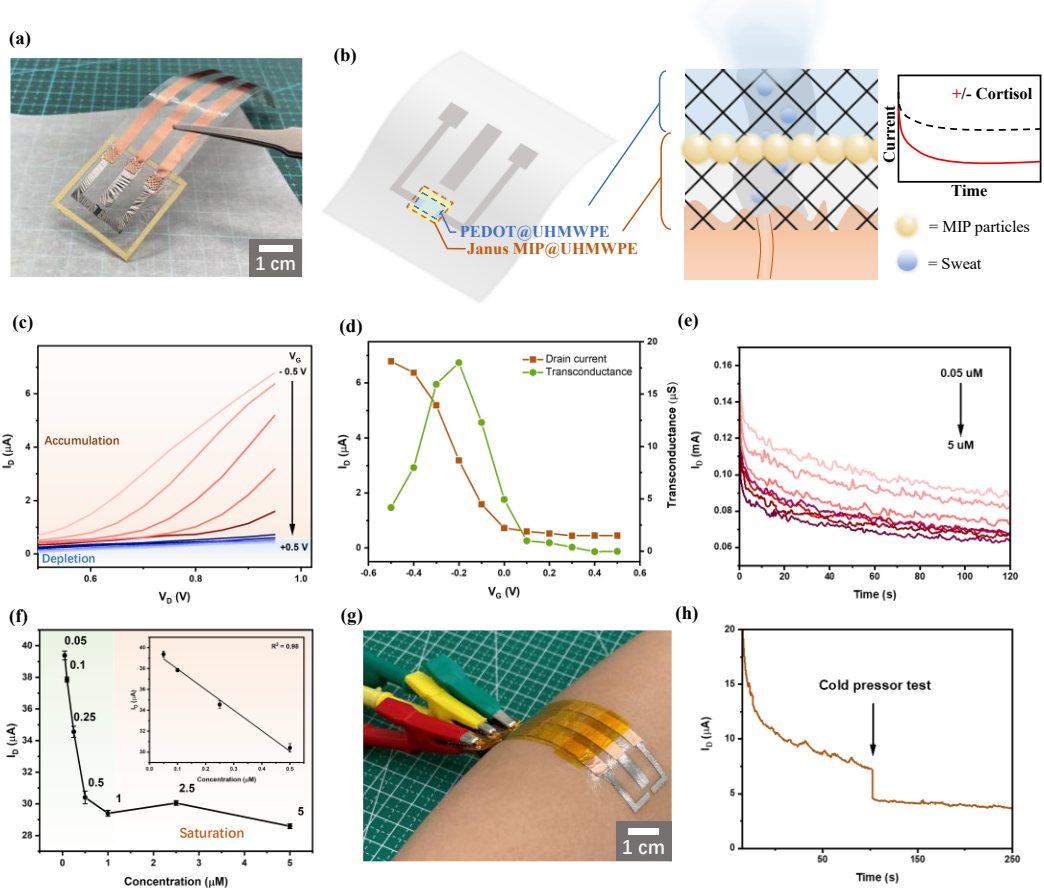

**Figure 4. UHMWPE-based sweat sensor.** (a) A photograph of UHMWPE-based sweat sensor. (b) a schematic illustration of sensor design and biomarker transport. (c) ID-VD curve of UHMWPE-based sweat sensor with different VG. (d) ID-VG curve and transconductance of UHMWPE-based sweat sensor. (e) Ex situ sensitivity test toward cortisol. (f) Ex situ calibration curve. (g) A photograph of UHMWPE-based sweat sensor conformed on the forearm of a volunteer. (h) On-body sensitivity test under CPT.